\documentclass[aps,prl,twocolumn,showpacs]{revtex4}
\usepackage{graphicx}
\usepackage{amsmath}
\usepackage{amssymb}

\begin{document}
\title{Broken-Symmetry States of Dirac Fermions in Graphene with A Partially Filled
High Landau Level}

\author{Hao Wang$^{1}$, D. N. Sheng$^1$, L. Sheng$^{2}$, and
F. D. M. Haldane$^3$}
\affiliation{ $^1$Department of Physics and
Astronomy, California State
University, Northridge, California 91330, USA\\
$^2$National Laboratory of Solid State Microstructures and
Department of Physics, Nanjing University, Nanjing 210093, P.
R. China\\
$^3$Department of Physics, Princeton University, Princeton, NJ
08544, USA}

%\date{\today}

\begin{abstract}
We report on numerical study of the Dirac fermions in partially
filled $N=3$ Landau level (LL) in graphene. At half-filling, the
equal-time density-density correlation function displays sharp peaks
at nonzero wavevectors $\pm {\bf q^{*}}$. Finite-size scaling shows
that the peak value grows with electron number and diverges in the
thermodynamic limit, which suggests an instability toward a charge
density wave. A symmetry broken stripe phase is  formed at large
system size limit, which is robust against purturbation from
disorder scattering. Such a quantum phase is experimentally
observable through transport measurements. Associated with the
special wavefunctions of the Dirac LL, both stripe and bubble phases
become possible candidates for the ground state of the Dirac
fermions in graphene with lower filling factors in the $N=3$ LL.
\end{abstract}

%73.43.-f Quantum Hall effects
%71.45.Lr Charge-density-wave systems (see also 75.30.Fv Spin-density waves)
%71.10.-w Theories and models of many-electron systems
%73.22.Gk Broken symmetry phases
%71.10.Pm Fermions in reduced dimensions
\maketitle

Recently, successful fabrication of single-atomic-layer-thick films
of graphite~\cite{G0}, called graphene, has triggered intensive
research activities to understand the novel
properties~\cite{G1,G2,G3,T0,G4} of these new two-dimensional (2D)
electron systems (2DES's). Different from the conventional 2D
electrons with the quadratic dispersion relation, the low-energy
electron excitations in graphene have a linear (relativistic)
dispersion relation, which can be described by a massless Dirac
equation~\cite{T1,T2,T3}. The Dirac-fermion-like nature of the
electrons has manifested itself evidently in the unconventional
quantization pattern of integer quantum Hall effect
(IQHE)~\cite{G1,G2,G3,T0,G4,haldaneh,donnah,odd2,odd3}. The
possibilities to observe the fractional quantum Hall effect in
graphene in the partially filled $N=0$ and $N=1$ LLs and a
pseudospin ferromagnetic state in the $N=1$ LL have been predicted
in some recent theoretical works~\cite{odd2,odd3,FQHE,lisheng},
which however have not been experimentally observed yet. Thus the
role of impurity scattering seems crucially important, which may
obscure the observation of these quantum phases for currently
available sample mobility. At even higher LLs, there exist many
studies on the conventional 2DES's, where a variety of quantum
phases have been predicted~\cite{first,mac,rhy,rhy1} and discovered
experimentally~\cite{experi}, varing from stripe phase to reentrant
QHE. On the other hand, for the Dirac fermions with partially filled
high LLs in graphene, investigations are only carried out based on
the Hartree-Fock mean-field approximation~\cite{hf}, which suggested
that stripe and bubble phases are possible. It remains an open issue
whether such quantum phases can survive quantum fluctuations and
disorder scattering. Study of the quantum phases of Dirac fermions
in high LLs is highly valuable for elucidating the key role of the
Coulomb interaction and the disorder effect in graphene.

In this Letter, we investigate the low-energy states of the Dirac
fermions in graphene with partially-filled $N$=3 Dirac LL using
Lanczos method for finite-size systems with torus geometry and upto
$N_e=16$ electrons in the LL~\cite{haldane1}. For a pure system at
half-filling, we find a large number of nearly degenerate low-energy
states, which are equally distanced from each other by a
characteristic wavevector ${\bf q^{*}}$ in momentum space. The
equal-time density-density correlation function shows strong and
sharp peaks at $\pm {\bf q^{*}}$, indicating the charge density wave
(CDW) instability. Finite-size scaling shows that the peak value
grows with electron number and becomes divergent in the
thermodynamic limit. Thus a symmetry broken stripe phase will be
formed at large system size limit, which is robust against moderate
disorder scattering with a critical disorder strength comparable
with that for the nonrelativistic 2DES's~\cite{donnas}. Due to the
special single-particle wavefunctions of the Dirac LL, both stripe
and bubble phases are possible candidates for the ground state with
lower filling factors in the $N=3$ LL based on exact calculations.
We further discuss the transport anisotropy of the stripe phase,
which can be used to experimentally detect this quantum phase.

We consider a 2DES in an $L_x\times L_y$ rectangular cell of
graphene under a perpendicular magnetic field. The magnetic length
$\ell$ is taken to be the unit of length. The total number of flux
quanta $N_{\phi}=L_x L_y/2\pi$ is chosen to be an integer. Periodic
boundary conditions are imposed in both $x$ and $y$
directions~\cite{rhy}. The magnetic field is assumed to be strong
enough so that the spin degeneracy of the LLs is lifted, and all the
LLs are well separated from each other. One can thus project the
system Hamiltonian into the topmost, partially filled, $N$-th
LL~\cite{rhy}. The projected Hamiltonian, which contains the Coulomb
interaction and disorder potential, has the form:
\begin{eqnarray}
% H_{c}&=&\sum_{i<j}\sum_{\textbf{q}}e^{-q^2/2}\frac {F_N^2(q)} {2\pi N_{\phi}}
% V(q)e^{i\textbf{q}\cdot(\textbf{R}_i-\textbf{R}_j)}\nonumber\\
 H_{c}&=&\sum_{i<j}\sum_{\mathbf{q}}e^{-q^2/2}[F_N(q)]^2
 V(q)e^{i\mathbf{q}\cdot(\mathbf{R}_i-\mathbf{R}_j)}/2\pi N_{\phi}\nonumber\\
 &+&\sum_{i}\sum_{\mathbf{q}}e^{-q^2/4}F_N(q)
 V_{\rm imp}(q)e^{i\mathbf{q}\cdot\mathbf{R}_i}\ ,
\end{eqnarray}
where $\mathbf{R}_i$ is the guiding center coordinate (GCC) of the
$i$-th electron, and $V(q)=2\pi e^2/\epsilon q$ is the Fourier
transform of the Coulomb interaction. The wavevector ${\bf q}$ takes
discrete values that are compatible with the geometry of the system.
Noticing that the electron wavefunction in a relativistic Dirac LL
of $N$ is a mixture of two wavefunctions in two different
nonrelativistic LLs of $N$ and $N-1$, we can write the form factor
$F_N(q)$ in Eq.\ (1) as~\cite{odd2}
\begin{equation}
F_N(q)=\frac{1}{2}[L_N(q^2/2)+L_{N-1}(q^2/2)]\ ,
\end{equation}
where $L_N(x)$ is the Laguerre polynomial. The disorder potential is
generated according to the correlation relation in $q$-space
$\langle V_{\rm imp}(q)V_{\rm imp}(-q')\rangle=\frac {W^2} {L_xL_y}
\delta _{q,-q'}$, which corresponds to $\langle V_{\rm imp}({\bf
r})V_{\rm imp}({\bf r'})\rangle=W^2 \delta (\bf {r-r'})$ in real
space with $W$ as the strength of disorder in units of
$e^2/\epsilon\ell$.

%%%% FIGURE 1 %%%%
\begin{figure}
\centerline{\includegraphics[width=3.0 in]{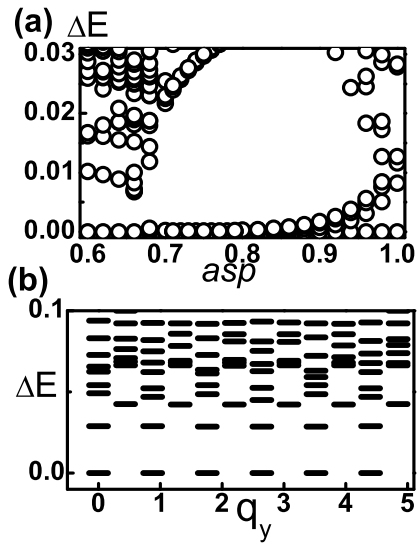}} \caption{
Low-energy spectrum for the half-filled $N=3$ LL with $N_{e}=12$ in
a clean system. (a) Energy levels versus aspect ratio $asp$, where
the eigenenergies are measured in units of $e^{2}/\epsilon\ell$ with
respect to the ground state. (b) Energy levels versus the $y$
component of wavevector at aspect ratio $asp=0.74$. The six nearly
degenerate lowest-energy states are equally separated by a
wavevector $\mathbf{q}^{*}=(0, 4\pi/L_y)=(0,0.88)$.}
\end{figure}

We compute exactly the low-energy spectrum and wavefunctions using
the Lanczos diagonalization method. We find that
pseudospin-polarized states have lower energies than unpolarized
states. In Fig.\ 1a, the low-energy spectrum for the $N=3$ LL with
filling factor $\nu=1/2$ and $N_{e}=12$ is shown as a function of
the aspect ratio $asp=L_{x}/L_{y}$. Here, the filling factor is
defined as $\nu=N_{e}/N_{\phi}$ with $N_{e}$ the electron number in
the LL. A generic feature of the spectrum is the existence of a
number of nearly degenerate low-energy states well separated from
higher-energy states by a gap, in a wide range of aspect ratio
$0.7<asp<0.9$. We will call these states the ground-state manifold.
In Fig.\ 1b, we plot the low-energy spectrum as a function of
wavevector $q_y$ at the optimized aspect ratio $asp=0.74$, where the
energy broadening of the manifold is minimized. All the states in
the manifold are equally separated from each other by a
characteristic wavevector $\mathbf{q}^{*}=(0, q^{*}_{y})$ with
$q^{*}_y=4\pi/L_y$. This feature is similar to that observed in the
spectra of the nonrelativistic 2DES's with half-filled high
LLs~\cite{rhy}, which indicates an instability towards a
unidirectional CDW or stripe phase.

%%%% FIGURE 2 %%%%
\begin{figure}
\centerline{\includegraphics[width=2.8in]{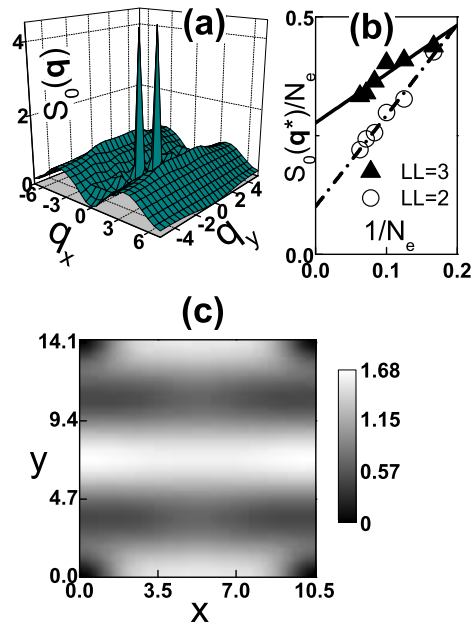}} \caption{(Color
online) Correlation functions of the ground state for the
half-filled $N=3$ LL in a clean system. (a) Static density-density
correlation function $S_0(\mathbf{q})$ with $N_e=12$ and $asp=0.74$.
(b) Peak value of $S_0(\mathbf{q})$ normalized by $N_{e}$ as a
function of $1/N_{e}$ for $N$=2 and 3 LLs. (c) Ground-state pair
correlation function $g(\mathbf{r})$ in guiding center coordinates
with $N_e=12$ and $asp=0.74$.}
\end{figure}

We next turn to the LL projected equal-time density-density
correlation function $S_0(\bf q)$ of the ground state~\cite{rhy}. In
Fig.\ 2a, $S_{0}({\bf q})$ for the half-filled $N=3$ LL with
$N_e=12$ and aspect ratio $asp=0.74$ is shown in a 3D plot. We see
that there are two sharp and strong peaks at $\pm
\mathbf{q}^{*}=(0,\pm 0.88)$ with the peak value 4.37, while
$S_0(q)$ is about $0.5$ away from the peaks. The presence of the
peaks in $S_{0}({\bf q})$
at  $\mathbf{q}^{*}$     suggests strong
density correlation at the
ordering wavevector, which is consistent with the characteristic
feature of the low-energy spectrum. We have also examined
 $S_{0}({\bf q})$ at a number
of other aspect ratios in the range of $0.7<asp<0.9$, which all show
sharp peaks  but with slightly reduced peak values.
The ratio $\frac
{S_0({\bf q}^{*})}{N_e} $ for different electron numbers from
$N_e=6$ upto $N_e=16$ is shown as a function of $1/N_e$  in Fig.\
2b, at the corresponding optimized aspect ratios. The ratio
$\frac{S_0({\bf q}^{*})}{N_e}$ extrapolates linearly to a finite
value 0.28 as $N_{e}\rightarrow\infty$, which is proportional to the
relative density  modulation in the symmetry broken state.
Therefore, the instability towards a unidirectional CDW phase in
large systems is established by finite-size scaling. We have also
studied the $N$=2 LL, where both the value of $S_0(\bf{q}^*)$ and
the value of $\frac{S_0({\bf q}^{*})}{N_e}$ at
$N_{e}\rightarrow\infty$ are smaller than those in the $N=3$ LL case
suggesting the $N=3$ LL to be the better candidate for observing the
CDW phases in graphene.

%%% FIGURE 3 %%%%%%%
\begin{figure}
\centerline{\includegraphics[width=3.2in]{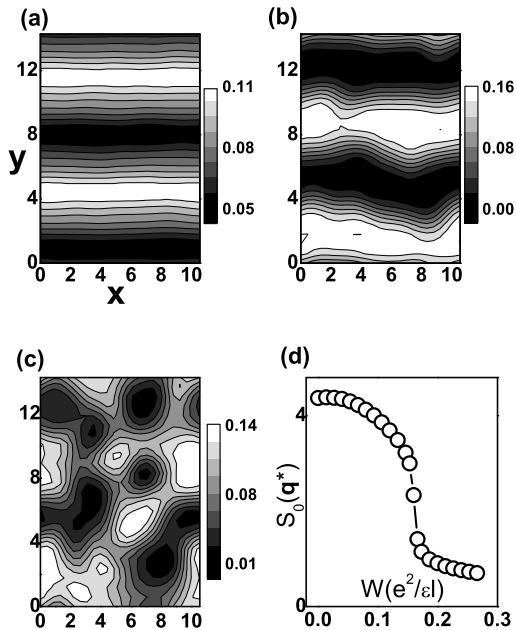}}\caption{Disorder
effect on the ground state of the half-filled $N=3$ LL with $N_e=12$
and $asp=0.74$. Projected electron density $\rho(\mathbf{r})$ at
disorder strengths (a) $W=0.01$, (b) $W=0.10$, and (c) $W=0.20$. (d)
Peak value of $S_0(\mathbf{q})$ versus disorder strength. The
transition occurs around $W=0.16$.}
\end{figure}

The CDW order of the above ground state in the real space is studied
using the LL projected pair correlation function $g(r)$
~\cite{yoshioka}. In Fig.\ 2c, we plot the ground-state pair
correlation function in GCC for the half-filled $N=3$ LL with
$N_e=12$ and $asp=0.74$. It shows clearly that there are two stripes
inside the unit cell along the $x$ direction. The mean separation
between the two stripes is related to the ordering wavevector
$q^{*}_y$ through $D_s=2\pi/q^{*}_y=7.1$, which happens to be a half
of $L_y$ for the present system. By examining systems with $N_{e}$
ranging from $6$ upto $16$ at their optimized aspect ratios, we find
that the ordering wavevector $q_y^{*}$ changes slightly with $N_{e}$
between $q_y^{*}=0.88$ and $0.96$, and correspondingly $D_s$ varies
between 6.5 and 7.1 magnetic lengths, which are comparable to the
ones for the nonrelativistic 2DES's~\cite{rhy}.

The above calculated correlation functions establish a stripe phase
at the thermodynamic limit, in agreement with the Hartree-Fock
result~\cite{hf}. We also note that in a clean finite-size system
($W=0$),  the ground state wavefunction is invariant under the
magnetic translation of  guiding centers of the cyclotron orbits,
while the symmetry broken stripe phase can become the true ground
state when a relatively weak disorder is turned on~\cite{donnas}. In
the presence of such random disorder, we can directly probe the CDW
phase using the LL projected local electron density $\rho({\bf
r})$~\cite{donnas}. For a very weak disorder strength, the typical
behavior of $\rho({\bf r})$ is shown in Fig.\ 3a, where $W=0.01$.
Two nearly perfect stripes are formed along the $x$ direction with
essentially no density modulations along the stripes. At a
intermediate disorder strength $W=0.1$, there are still two complete
stripes, though the shape of the stripes exhibits pronounced
variations along the stripe direction. Thus we conclude that for
weak to moderate disorder strength, the ground state is in the
quantum Hall stripe phase. With further increasing disorder
strength, as shown in Fig.\ 3c for $W=0.2$, the stripe density
profile becomes raptured and riddled with defects, indicating that
the long-range stripe order is lost. The remaining short-range
stripes become randomly orientated. In the relatively weak disorder
region $0<W<0.1$, the whole structure of the density-density
correlation function $S_0(\mathbf{q})$ is found to be similar to the
clean system case, indicating the robustness of the CDW order. In
Fig.\ 3d, we plot the peak value $S_0(\mathbf{q}^*)$ as a function
of $W$. The peak value $S_0(\mathbf{q}^*)$ remains nearly constant
for relatively weak disorder $W<0.1$. It starts to drop quickly
around $W=0.13$ and becomes comparable with the background value 0.5
at $W>0.16$. Thus $W=0.16$ is determined as the critical disorder
strength, where the long-range CDW order becomes unstable.

The transport property of the above anisotropic CDW state is studied
by calculating the Thouless energy $\Delta E _{\tau \tau}$ ($\tau=x$
or $y$), i.e., the ground state energy difference due to a change of
the boundary condition in the $\tau$ direction from periodic to
antiperiodic, which is related to the longitudinal conductance in
the $x$ or $y$ direction. We find that for a weak disorder, e.g.,
$W=0.02$, $\Delta E_{yy}$ is about 50 times smaller than $\Delta
E_{xx}$, suggesting a large transport anisotropy associated with the
quantum Hall stripe phase, which can be used to identify the stripe
phase experimentally.

%%%% FIGURE 4 %%%%
\begin{figure}
\centerline{\includegraphics[width=2.8 in]{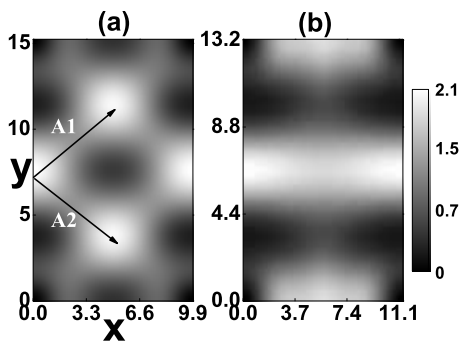}}
\caption{Ground-state pair correlation functions for the $N=3$ LL
for filling factor $\nu=1/3$ and electron number $N_{e}=8$ at the
aspect ratios (a) $asp=0.65$ and (b) $asp=0.86$.}
\end{figure}

We further study the ground state at lower filling factors in the
$N=3$ LL. For $\nu=1/3$ and $N_e=8$, the low-energy states are
found to become nearly degenerate in two different aspect ratio
regions. The ground-state pair correlation functions corresponding
to these two regions are plotted in Fig.\ 4(a) and Fig.\ 4(b),
where the aspect ratio is taken to be $asp=0.65$ and $asp=0.86$,
respectively. In the region around $asp=0.65$, the ground state is
a bubble-like CDW phase~\cite{rhy1}. The bubbles distribute in a
superlattice structure with the lattice vectors $\mathbf{A1}$ and
$\mathbf{A2}$. For $asp=0.65$, we have $A1=A2=6.24$, and the angle
between the two lattice vectors is $\theta=75^{\mathrm{o}}$.
Within the unit cell, there exist four bubbles and each bubble
contains two electrons. In the region around $asp=0.86$, the
ground state is a stripe phase, and the stripe separation is about
$D_s=6.62$ for $asp=0.86$. These results indicate that both the
two-electron bubble phase and the stripe phase can be separately
realized, depending on the geometry of the system. This is
different from the case for nonrelativistic 2DES's, where usually
only a single CDW phase exists at a given filling factor in a
given LL~\cite{rhy1}. More competing phases in the Dirac LL may be
explained as a result of the special structure of the Dirac LL,
which is essentially a combination of the $N=2$ and $N=3$
nonrelativistic LLs.

In summary, we have shown that various CDW phases can be realized in
the partially-filled $N=3$ LL in graphene, the character of which
can be explained from the mixture property of the electron
wavefunctions of the Dirac LL. The CDW phases are robust against
moderate disorder scattering. The unidirectional CDW stripe phase at
half-filling shows large transport anisotropy, which can be measured
experimentally.

\textbf{Acknowledgment:} This work is supported by the DOE grant
DE-FG02-06ER46305,  the NSF grants DMR-0605696 (DNS) and DMR-0611562
(HW, DNS), the National Basic Research Program of China 2007CB925104
(LS), the NSF under MRSEC grant DMR-0213706 at the Princeton Center
for Complex Materials (FDMH), and the support from KITP (through NSF
grant PHY05-51464) and KITPC.

\end{document}